\documentclass[11pt,a4paper]{llncs}
\usepackage{amsmath}
\usepackage{amssymb}
\usepackage{graphicx}
\usepackage{marvosym}
\usepackage{url}
\usepackage{fancyhdr}

\usepackage[T1]{fontenc}
\usepackage{adjustbox}
\usepackage{booktabs}
\usepackage{graphicx}
\usepackage{makecell}
\usepackage{pifont}
\usepackage{enumitem}
\usepackage{hyperref}
\usepackage{url}
\usepackage[utf8]{inputenc}
\usepackage[T1]{fontenc} 
\usepackage{amsmath}

\usepackage{booktabs} 
\usepackage{tcolorbox}
\usepackage{caption}
\usepackage{pifont}
\usepackage{adjustbox}
\usepackage{makecell}
\usepackage{enumitem}

\usepackage{geometry}
\newcommand{\keywords}[1]{\par\addvspace\baselineskip
\noindent\keywordname\enspace\ignorespaces#1}

\fancypagestyle{firstpage}{\fancyhf{}
}

\begin{document}


\title{\LARGE{Deep Learning Aided Software Vulnerability Detection: A Survey}}

%
%

\author{}

\author{\large{Md Nizam Uddin  \and Yihe Zhang \and Xiali Hei}}
\institute{\large{University of Louisiana at Lafayette, Lafayette LA 70503, USA}}

\maketitle

\thispagestyle{firstpage}

\begin{abstract}
The pervasive nature of software vulnerabilities has emerged as a primary factor for the surge in cyberattacks.
Traditional vulnerability detection methods, including rule-based, signature-based, manual review, static, and dynamic analysis, often exhibit limitations when encountering increasingly complex systems and a fast-evolving attack landscape.
Deep learning (DL) methods excel at automatically learning and identifying complex patterns in code, enabling more effective detection of emerging vulnerabilities.
%
%
This survey analyzes 34 relevant studies from high-impact journals and conferences between 2017 and 2024.
This survey introduces the conceptual framework ``Vulnerability Detection Lifecycle'' for the first time to systematically analyze and compare various DL-based vulnerability detection methods and unify them into the same analysis perspective. 
The framework includes six phases: (1) Dataset Construction, (2) Vulnerability Granularity Definition, (3) Code Representation, (4) Model Design, (5) Model Performance Evaluation, and (6) Real-world Project Implementation.
For each phase of the framework, we identify and explore key issues through in-depth analysis of existing research while also highlighting challenges that remain inadequately addressed.
This survey provides guidelines for future software vulnerability detection, facilitating further implementation of deep learning techniques applications in this field.

%
%

\end{abstract}


\section{Introduction}
Cyberattacks, which are reported to occur approximately every 39 seconds, causing an average loss of \$2.6 million per incident and impacting over 800,000 victims annually, have imposed a significant financial burden on companies and governments~\cite{explodingtopics2023}. 
These attacks involve attackers gaining unauthorized access or exploiting malicious actions through software vulnerabilities (\textit{e.g.}, weaknesses and flow) in a computer system. 
The vulnerabilities that arise from inadvertent software practices contribute to programmer cognition failures during design or implementation and are often introduced during the development process.   
Moreover, with the abundance of software reuse in this era of Open Source Software (OSS) development, the dissemination of vulnerabilities has presented a tougher challenge to security experts than ever~\cite{gkortzis2021software}.

A multitude of sophisticated methodologies have been proposed to address the pervasive issue of software vulnerability. Static analysis~\cite{cousot1977abstract} involves the meticulous examination of source code or compiled code to identify potential vulnerabilities. Dynamic analysis~\cite{coleman1972program} entails executing the program and meticulously analyzing its runtime behavior to detect vulnerabilities. Furthermore, Fuzz testing~\cite{miller1989fuzz} involves generating random inputs to a program to uncover vulnerabilities that can be triggered by unforeseen inputs.

Software vulnerabilities continue to be a serious hazard to computer systems even with the effectiveness of these solutions. This is caused by the complexity of software systems and the dynamic character of cyber-attacks~\cite{zhou2020ever}. Hence, the development of new, reliable techniques that need less human labor and produce high levels of accuracy in software vulnerability identification is required. In a number of fields, such as computer vision and natural language processing, deep learning (DL) has demonstrated notable advancements~\cite{Chen2023}. 
As a result, researchers are now interested in using DL methods to automatically and accurately identify vulnerabilities.

\begin{figure}[h]
\centering
\includegraphics[scale=0.7]{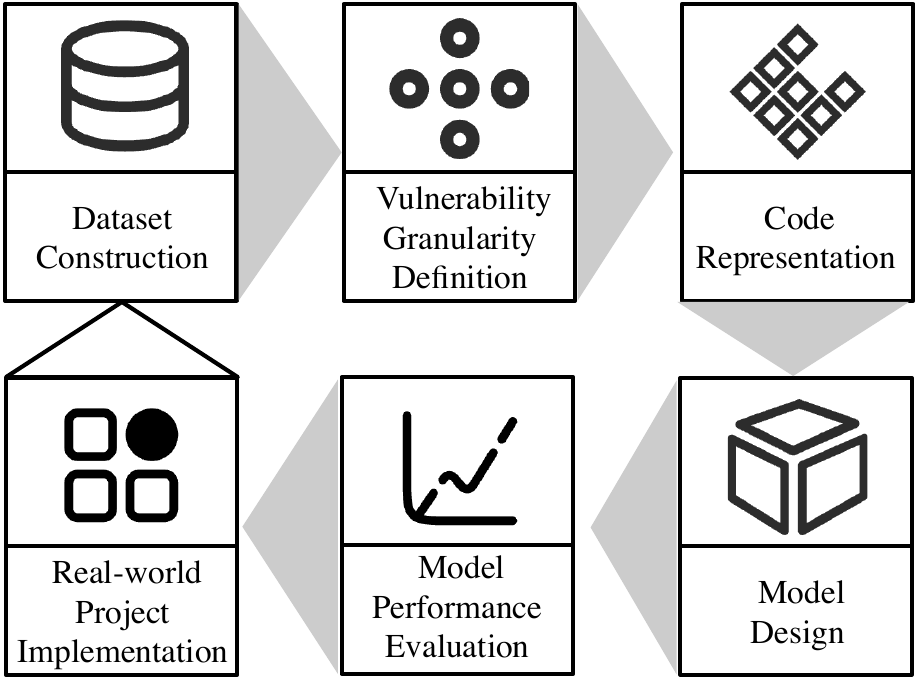}
\caption{Lifecycle of deep learning-based source code vulnerability detection.}
\label{Fig:Lifecycle}
\end{figure}

Each year, production software is exposed to tens of thousands of security flaws that can either be found internally in proprietary code or openly reported to the Common Vulnerabilities and Exposures (CVE) database~\cite{CVE}. These flaws frequently elude developers and code evaluators due to their subtle nature. 
However, there is a chance to gain insight into bug trends from the abundance of open-source code that is available for examination.

The two complementary methods, static analysis and dynamic analysis of source code, are used to find program vulnerabilities. 
Both methods, however, have the potential to yield erroneous positives and false negatives. 
On the other hand, Fuzzing is a method for finding software flaws by creating and feeding unexpected data. It requires a lot of resources and time, yet it can produce false positives or miss some types of vulnerabilities, consequently incurring high human intervention and demanding expertise to produce good results.

Deep learning techniques, with their prominent success in almost every aspect of scientific research, are no different in the field of source code vulnerability detection but with some weaknesses and challenges to overcome. The significance of this challenge has made it a prominent focus of recent studies. Accordingly, a comprehensive review has been conducted here, encompassing 34 studies published in esteemed journals and conference proceedings of IEEE, ACM, Springer, and Google Scholar between 2017 and 2024.  

With the proliferation of research in this field of study, no one has comprehensively carried out a systematic literature review that presents the life cycle of DL-based source code vulnerability detection. By making the following contribution, our study intends to fill up the gaps in existing review articles:  

\begin{itemize}[leftmargin=*, itemsep=0pt, labelsep=5pt]
    \renewcommand\labelitemi{\ensuremath{\bullet}}
    \item Firstly, we define the complete life cycle and core components of source code vulnerability detection using deep learning, as shown in Fig.~\ref{Fig:Lifecycle}.
    \item Secondly, we summarize core components by studying how the existing literature addressed each of the components and what their work methodology is within the component alongside their faced challenges.
    \item Finally, we discuss key challenges and pin down outstanding research gaps with potential future research direction.
\end{itemize}

The remainder of this paper is organized as follows: Section~\ref{sec:Research Method} provides an overview of the lifecycle of deep learning-based source code vulnerability detection and outlines our research methodology. Section~\ref{sec:components} presents a component-based discussion of the established lifecycle. Section~\ref{sec:Future Research} explores future research directions, and Section~\ref{sec:conclusion} concludes the paper.

\section{Research Method}
\label{sec:Research Method}

Source code vulnerability detection differs greatly from other domains in diverse ways where deep learning models are employed. The way source code is interpreted and represented in the model is one of them. An additional pivotal contrast pertains to the requirement for contextual comprehension of the source code. As vulnerabilities frequently hinge on particular programming constructions, models must comprehend deeper semantic links in addition to syntactic patterns.

Furthermore, models that can generalize across diverse and sometimes unknown code bases are required due to the complexity and variety of coding techniques across various languages and development environments. This section provides an overview of the elements of the DL-based source code vulnerability detection lifecycle alongside our research methodology. These components are illustrated in Fig.~\ref{Fig:Lifecycle}, including Dataset Construction, Vulnerability Granularity Definition, Code Representation, Model Selection, Model Performance Evaluation, and Real Project Implementation.

\subsection{Research Strategy}
In order to achieve the primary objectives of this study, we have adopted a comprehensive research strategy that encompasses the following steps:

\begin{enumerate}[leftmargin=*, itemsep=0pt, labelsep=5pt]
    \item \textbf{Define the Life Cycle of DL-Based Source Code Vulnerability Detection:} We meticulously outlined the various phases involved in the application of deep learning techniques to detect vulnerabilities in source code.
    
    \item \textbf{Select  Groundbreaking and Impactful Studies:} We identified and selected seminal and highly impactful studies due to their influence on subsequent research from the past seven years in the domain of DL-based vulnerability detection.
    
    
    \item \textbf{Analyze Methodologies and Challenges:} We examined the methodologies employed by the selected studies within each component of the life cycle, highlighting the challenges encountered. 
    
    \item \textbf{Identify Research Gaps:} We pinpointed existing research gaps through existing challenges in the field and providing a comprehensive overview of current and future directions. 
\end{enumerate}
This structured approach ensures a thorough investigation and critical analysis of the current state of research, ultimately contributing to the advancement of knowledge in this field.

\subsection{Selection of Studies}

To focus on recent studies from 2017 to the present, papers from IEEE, ACM, Springer, and Google Scholar were meticulously reviewed. The initial pool of 94 studies was refined to 34 after accounting for duplicates across databases and assessing their relevance. Studies that did not specifically address deep learning methods for vulnerability detection, including those focused on general machine learning techniques, were excluded. In contrast, studies explicitly employing deep learning methods for vulnerability detection were prioritized, with a particular emphasis on seminal and highly impactful research.

To determine seminal and highly impactful research in deep learning-based vulnerability detection, we adopt a three-pronged selection approach based on:

\begin{itemize}
\item \textbf{Novelty} – The study introduces a groundbreaking concept, methodology, or dataset that significantly advances the field.
\item \textbf{Citation Count} – The paper has garnered substantial citations, indicating its influence and recognition within the research community. The citation counts of our selected studies range from 127~\cite{chen2023diversevul} to 1,216~\cite{Li2018VulDeePecker}, underscoring their impact and widespread adoption in the field. 
\item \textbf{Foundational Impact} – The study serves as a cornerstone for subsequent research, meaning later works explicitly build upon, extend, or benchmark against it. This criterion captures the generational influence of research, identifying papers that have shaped the trajectory of deep learning in vulnerability detection.
\end{itemize}

Our rigorous selection process ensured that only the most pertinent and influential works were considered, providing a robust foundation for our analysis. The resulting selection highlights key advancements and trends in the application of deep learning to vulnerability detection. These selected papers collectively excels our survey work to represent a comprehensive overview of the most significant contributions to deep learning-based vulnerability detection in recent years and draw the future directions of this field.

\section{Component Based Discussion}
\label{sec:components}

\subsection{Dataset Construction}

The quality of the dataset is paramount to the success of DL models in vulnerability detection, as it significantly impacts the model's ability to learn and effectively generalize the vulnerability pattern ~\cite{Kotsiantis2006}. 
A well-constructed dataset enables the model to accurately capture underlying patterns and relationships within the data, thereby reducing the risk of over-fitting/under-fitting and enhancing predictive accuracy.

\begin{table*}[t] 
  \caption{Overview of most prominent Vulnerability Datasets}
  \label{tab:overview_vulnerability_datasets}
  \centering
  \begin{adjustbox}{max width=\textwidth} 
    \begin{tabular}{lccccccc}
      \toprule
      \textbf{Dataset} & \textbf{\makecell{Source Code\\Language}} & \textbf{Type} & \textbf{Source} & \textbf{Status} & \textbf{\makecell{Vuln. \\ Samples}} & \textbf{\makecell{Non-Vuln. \\ Samples}} & \textbf{Gran.} \\
      \midrule
      Juliet~\cite{nist2018juliet} & \makecell{C/C++,\\ Java} & Synth. & \makecell{Common Vulnerability \\ Pattern} & \ding{51} & 89,000 & 89,000 & $Level\;1$ \\
      Draper~\cite{Russell2018Draper} & C/C++ & \makecell{Synth.\& \\ Real} & \makecell{Open Source Projects \\ (OSP)} & \ding{51} & 82,324 & 1,192,041 & $Level\;2$ \\
      DeepWukong~\cite{cheng2021deepwokung} & C/C++ & Real & SARD \& 2 OSPs & \ding{51} & 156,195 & 409,262 & $Level\;3$ \\
      SySeVR~\cite{Li2021SySeVR} & C/C++ & Real & SARD \& NVD & \ding{51} & 56,395 & 364,232 & $Level\;3$ \\
      FUNDED~\cite{wang2020FUNDED} & C/C++ & Real & SARD \& NVD \&  OSP & \ding{51} & 75,474 & 75,474 & $Level\;2$ \\
      REVEAL~\cite{chakraborty2020reveal} & C/C++ & Real & OSP & \ding{51} & 1,664 & 16,505 & $Level\;2$ \\
      Devign~\cite{zhou2019devign} & C/C++ & Real & OSP & \ding{108} & 10,067 & 12,294 & $Level\;2$ \\
      D2A~\cite{zheng2021d2a} & C/C++ & Real & OSP & \ding{51} & 18,653 & 1,295,623 & $Level\;3$ \\
      Vuldeepecker~\cite{Li2018VulDeePecker} & C/C++ & Real & NVD \& SARD & \ding{51} & 17,725 & 43,913 & $Level\;3$ \\
      µVulDeePecker~\cite{Zou2019muvuldeppecker} & C/C++ & Real & NVD \& SARD & \ding{51} & 43,119 & 138,522 & $Level\;3$ \\
      DiverseVul~\cite{hussain2024vulnerability} & C/C++ & Real & OSP & \ding{51} & 18,945 & 330,492 & $Level\;2$ \\
      Hussain et al.~\cite{chen2023diversevul} & Java & Real & SARD & \ding{51} & 29,258 & 12,954 & $Level\;4$ \\
      \bottomrule
    \end{tabular}
  \end{adjustbox}
  \raggedright \textbf{\small \ding{51}: Publicly Available, \ding{108}: Partially Available}
  \vspace{-1em}
\end{table*}


National Vulnerability Database (NVD) ~\cite{NVD} and the Software Assurance Reference Dataset (SARD) ~\cite{NIST_SARD} have been a prominent source of data for vulnerable code due to reliability and credibility. NVD's authority stems from its affiliation with the National Institute of Standards and Technology (NIST) \cite{nist_website} in the United States, ensuring meticulous curation and adherence to rigorous standards. On the other hand, SARD's reliability lies in its role as a curated repository of diverse and well-documented test cases, meticulously designed to evaluate the efficacy of software quality assurance tools. Several state-of-the-art works have extensively relied on these two prominent sources for dataset preparation namely VulDeePecker~\cite{Li2018VulDeePecker} SySeVR~\cite{Li2021}, VulHunter~\cite{Guo2020}, {\textmu}Vuldeepecker~\cite{Zou2019muvuldeppecker}.
Table ~\ref{tab:overview_vulnerability_datasets} presents the overview of the prominent vulnerability datasets.

\begin{figure}[t]
\centering
\includegraphics[scale=0.45]{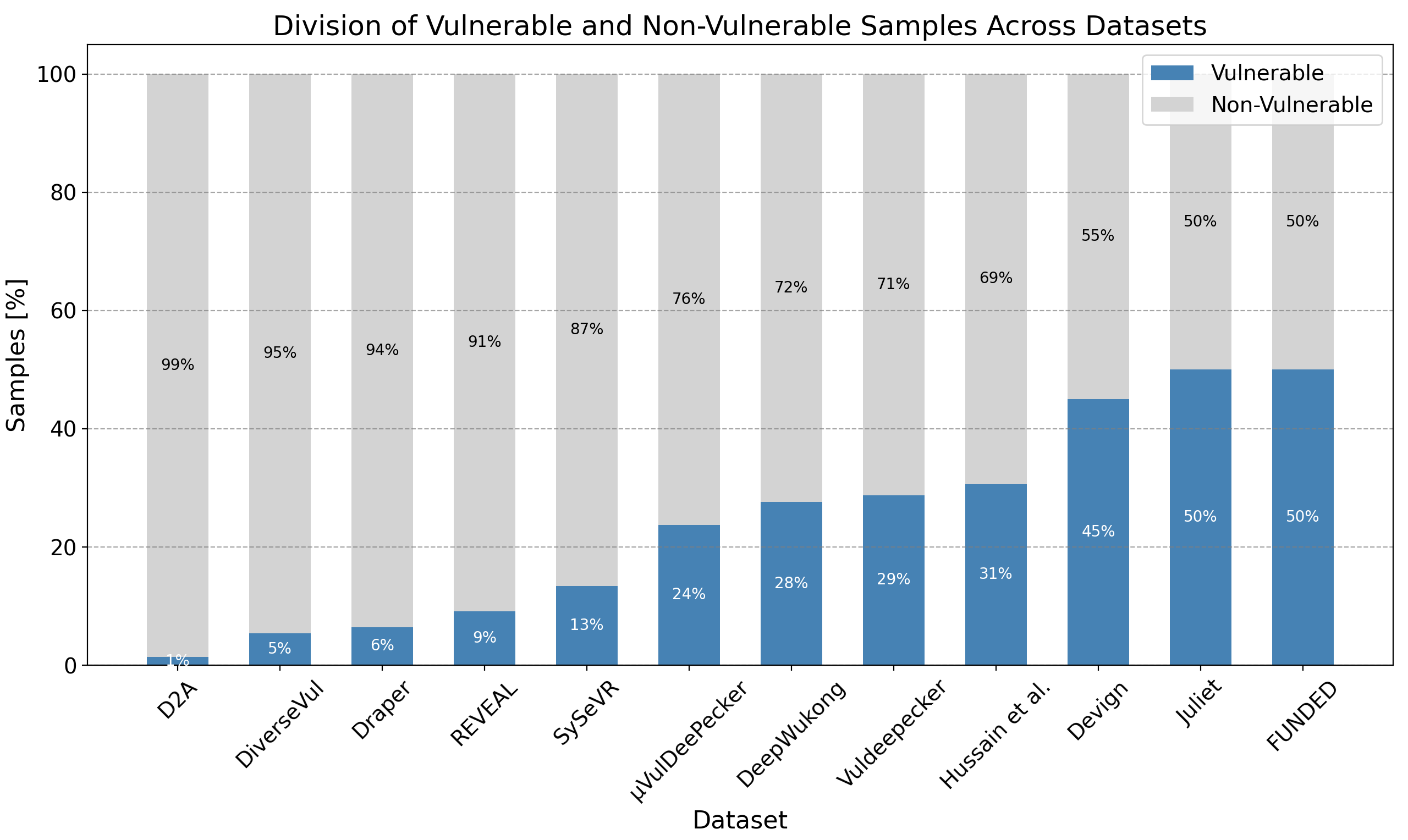}
\caption{Data imbalance across studied datasets.}
\label{fig:imbalance}
\vspace{-1em}
\end{figure}

\tcbset{colframe=black, colback=white, boxrule=0.4mm, arc=4mm, auto outer arc, width=\textwidth}
\begin{tcolorbox}[width=\linewidth]
\textbf{Challenge:} \textit{A large body of closely examined literature reveals that duplication and dataset imbalance (Fig.~\ref{fig:imbalance}) are widespread concerns.}
\end{tcolorbox}

{\textmu}Vuldeepecker~\cite{Zou2019muvuldeppecker} dataset  contains 181,641 pieces of code (called code gadgets, which are units for vulnerability detection) from 33,409 programs. Among them, 138,522 are non-vulnerable (i.e., not known to contain vulnerabilities)
and the other 43,119 are vulnerable and contain 40 types of Common Weakness Enumeration (CWEs)~\cite{cwe_website}.
Another type of dataset emerged according to predefined patterns that are called synthetic datasets. One of those is Juliet~\cite{NIST_SARD}  which is collection of test cases in C/C++ containing 64,099 test cases of 118 different CWEs. s-bAbI~\cite{Mikolov2018} contains syntactically valid C programs with non-trivial control flow with safe and unsafe buffer writes labeled at the line of code level, which is originally an extension of Facebook AI Research developed bAbI~\cite{Weston2015}.

\tcbset{colframe=black, colback=white, boxrule=0.4mm, arc=4mm, auto outer arc, width=\textwidth}
\begin{tcolorbox}[width=\linewidth]
\textbf{Challenge:} \textit{Synthetic and unrealistic source code are largely absent of the complexity and variability of actual software environments and many datasets are curated synthetically.}
\end{tcolorbox}

Static analyzers' and Synthetic datasets' labels are frequently imprecise in terms of accuracy and confidence.
Thus, the integrity of label assignments may be seriously jeopardized when vulnerability detection techniques are evaluated with these datasets~\cite{kang2022detecting}.
Conversely, real-world projects exhibit a broader spectrum of vulnerabilities. Open source projects(OSP) are a robust source of both vulnerable and non-vulnerable code examples through security patches, with platforms like GitHub and other public repositories providing excellent access for researchers.

\tcbset{colframe=black, colback=white, boxrule=0.4mm, arc=4mm, auto outer arc, width=\textwidth}
\begin{tcolorbox}[width=\linewidth]
\textbf{Challenge:} \textit{Despite the abundance of open-source projects (OSP), datasets are still limited to capturing only few top most severe software errors classified in CWE. This restriction poses a significant challenge as it may overlook less well-known vulnerabilities that can pose substantial threats to systems.}
\end{tcolorbox}

Russel et al.~\cite{Russell2018Draper} used Debian project while DeepWukong~\cite{cheng2021deepwokung} dataset is based on the combination of SARD and two real-world open-source c/c++ projects, lua and redis. Lin et al.~\cite{Lin2019} constructed a real-world vulnerability ground truth dataset containing manually labelled 1,471 vulnerable functions and 1,320 vulnerable files from nine open-source software projects. Fan et al.~\cite{Fan2020} proposed Big-Vul dataset that includes detailed CVE IDs, severity scores, summaries, and code changes, all meticulously curated and stored in CSV format.

\tcbset{colframe=black, colback=white, boxrule=0.4mm, arc=4mm, auto outer arc, width=\textwidth}
\begin{tcolorbox}[width=\linewidth]
\textbf{Challenge:} \textit{ The restrictive size of the datasets may not adequately represent the breadth and depth of real-world software systems and their vulnerabilities.}
\end{tcolorbox}

IBM’s research team produced D2A~\cite{zheng2021d2a}, which extracts samples through inter-process analysis, contrasting with conventional datasets with functions only. This approach facilitates the comprehensive preservation of bug types, specific locations, trace information, and analyzer output. Encompassing multiple prominent open-source software projects such as FFmpeg, httpd, Libav, LibTIFF, Nginx, and OpenSSL, the dataset provides a diverse and substantive foundation essential for rigorous research.


Meticulously compiled from various security issue websites by aggregating vulnerability reports, DiverseVul~\cite{chen2023diversevul} is an open-source vulnerability dataset encompassing 18,945 vulnerable functions and 330,492 non-vulnerable functions, derived from 7,514 commits, covering 150 different C/C++ CWEs.

\tcbset{colframe=black, colback=white, boxrule=0.4mm, arc=4mm, auto outer arc, width=\textwidth}
\begin{tcolorbox}[width=\linewidth]
\textbf{Challenge:} \textit{Existing datasets predominantly feature samples written in C/C++, with limited representation of programming languages such as Java, Python, and Swift. This disparity in language diversity poses a significant challenge in ensuring comprehensive coverage and applicability across various software ecosystems.}
\end{tcolorbox}

\subsection{Vulnerability Granularity Definition}

Vulnerabilities can manifest in various spaces within the source code. It is essential to identify the specific spaces that are particularly prone to these security issues. These critical spaces include but are not limited to, Files, Functions, Structures, Macros, and Variables. Each of these spaces can harbor distinct types of vulnerabilities, which necessitates careful examination and robust security measures to mitigate potential risks. In extensive source code repositories comprising thousands of lines of code, precise identification of vulnerabilities is imperative. 

\begin{figure}[t] 
    \centering
    \includegraphics[scale=0.7]{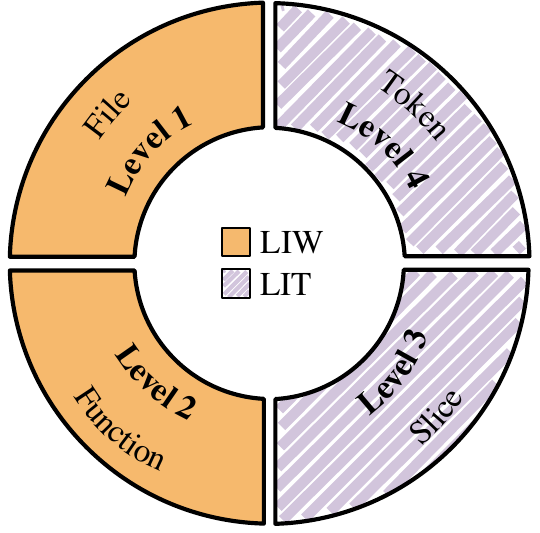} 
    \caption{Detection granularity with inherited issues.} 
    \label{fig:Granularity}
\end{figure}


We classify the granularity of vulnerability detection in existing research into four levels: 
file/program level ($Level\;1$), 
function level ($Level\;2$), 
slice level ($Level\;3$), 
and token level ($Level\;4$) 
Moreover, in Fig.~\ref{fig:Granularity} we depict the inherited problems for each level which are discussed subsequently.

Among studies, the most coarse-grained level of vulnerability detection is at the $Level\;1$. 
A list of 
works~\cite{Kalouptsoglou2020,niu2020deep,wang2016automatically} set their detection granularity at this level. 
The authors ~\cite{wang2016automatically} use the high-level semantic representations learned by the neural networks as defective features and conduct experiments on the dataset of open-source Java projects to identify $Level\;1$ vulnerabilities.
$Level\;2$ vulnerability detection is most frequently explored in this field, as evidenced by studies by
Vuddy~\cite{kim2017vuddy}, Mansur et al.~\cite{ahmadi2021finding}, Centris~\cite{woo2021centris}, MVP~\cite{xia2020mvp}, V1Scan~\cite{V1Scan}, Russel et al.~\cite{Russell2018Draper}, SySeVR~\cite{Li2021SySeVR}. 

Both $Level\;1$ and $Level\;2$ vulnerability detection suffers from the \textit{\textbf{``Lost in the Woods'' (LIW)}} problem, where a single marked vulnerable file or function may contain thousands of lines of code.
Vulnerabilities of external functional code are usually undetected and, therefore, require detection at a more granular level.

\tcbset{colframe=black, colback=white, boxrule=0.4mm, arc=4mm, auto outer arc, width=\textwidth}
\begin{tcolorbox}[width=\linewidth]
\textbf{Challenge:} \textit{LIW requires significant manual intervention from security experts to pinpoint the exact vulnerability.}
\end{tcolorbox}


The first work to employ DL to identify software vulnerabilities at the slice level ($Level\;3$) is from VulDeePecker~\cite{Li2018VulDeePecker}. 
This work creates a code gadget consisting of many program statements that are semantically related to each other but not necessarily consecutive. 
DeepWokung~\cite{cheng2021deepwokung} created code slice into a compact, low-dimensional representation, that preserves high-level programming logic, encompassing control and data flows, while maintaining the natural language context of the program.
The last line of works~\cite{islam2018towards,liu2020token,li2021enhancing,niu2020deep} presented token-level ($Level\;4$) detection granularity. It disassembles source code into discrete tokens, operators, identifiers, and keywords, providing extensive insights into code structure. Although this method may not fully capture intricate inter-token relationships, it serves as a foundational framework for code analysis. 
In their exploration of $Level\;4$ vulnerability detection, Islam et al.~\cite{islam2018towards} employed machine learning methodologies to dissect code segments. 
Furthermore, Li et al.~\cite{li2021enhancing} scrutinized the efficacy $Level\;4$ attributes for vulnerability detection, yielding promising outcomes in identifying security vulnerabilities.  

Loss of semantic integrity is one of the issues consistent with $Level\;3$ and $Level\;4$ vulnerability detection, which we name it \textit{\textbf{``Lost in Translation'' (LIT)}} problem. 
Although slicing and tokenization work well for capturing syntactic structures, they can miss subtle semantic linkages in the code, leading to inefficient feature learning and causing false positives or negatives.   
To improve the precision and dependability of detection systems, it is still imperative to strike a balance between syntactic slicing and semantic comprehension.

\tcbset{colframe=black, colback=white, boxrule=0.4mm, arc=4mm, auto outer arc, width=\textwidth}
\begin{tcolorbox}[width=\linewidth]
\textbf{Challenge:} \textit{LIT can seriously impair the effectiveness of the feature learning processes, thereby compromising the accuracy and reliability of vulnerability detection methods.}
\end{tcolorbox}

\subsection{Code Representation}

Once the granularity of detection has been established, the subsequent challenge involves presenting the code to autonomously learn important features. Using DL models to detect vulnerabilities accurately requires precise and meaningful source code representation. Code representations that capture syntactic patterns and semantic subtleties have been developed using a variety of methodologies. Existing studies on code representation techniques have been systematically categorized into five distinct approaches.


\subsubsection{Abstract Syntax Tree (AST)-Based Representation} captures the syntactic structure of source code using a tree of nodes, where each node represents a specific syntactic construct. This approach abstracts away low-level details while preserving essential syntactic information. The resulting AST provides a comprehensive view of the code's structure, including control flow and hierarchical relationships, enabling a more organized and systematic representation of the source code.
Although ASTs can incorporate specific semantic insights by utilizing techniques such as data flow analysis or type inference, their primary use is to capture the syntactic and structural aspects of the code~\cite{zhang2019novelAST}.
These require parsing the code into a tree structure, which is computationally manageable but can become complex with larger or more intricate code bases. 
The significance of hierarchical code representations was further emphasized by Ahmadi et al.~\cite{ahmadi2021finding}, who investigated AST-based characteristics for vulnerability identification in open-source projects. 
Furthermore, to illustrate the practicality of this method, Kalouptsoglou et al.~\cite{Kalouptsoglou2020} employed DL in conjunction with AST-based features to forecast cross-project vulnerabilities.

\subsubsection{Graph-Based Representation}
conceptualizes the source code as a graph structure where program items are represented adeptly, modeling syntax and semantics by incorporating different nodes and edges that depict not only the structure of code but also the interactions and dependencies, such as call graphs or program dependency graphs. 
This comprehensive approach enhances their ability to robustly model syntax and semantics ~\cite{nguyen2012graph,wang2020FUNDED,mi2023graph}.
It involves constructing and analyzing potentially large and complex graphs that represent various relationships and dependencies, demanding significant computational resources. 
Woo et al.~\cite{woo2021centris} highlighted scalability and efficiency by using graph-based characteristics for vulnerability identification in industrial code-bases.  
Additionally, Li et al. \cite{Li2021SySeVR} implemented graph-based representation by accommodating syntax and semantic information effectively. 


\subsubsection{Hybrid Representation}
comprehensively captures code syntax and semantics. 
They combine several levels of abstraction and thus inherit the computational demands of each, generally resulting in high complexity but superior analysis capabilities.
To improve the accuracy of vulnerability detection, Li et al.~\cite{li2020hybrid} presented a hybrid representation that combines token-level ($Level\;4$) features with higher-level semantic information. 
Furthermore, in comparison to traditional approaches, Cheng et al.~\cite{cheng2021deepwokung} achieved better performance with a hybrid strategy that combined AST-based characteristics with natural language processing techniques. 
In addition, Kim et al.~\cite{kim2017vuddy} investigated hybrid representations for vulnerability detection, blending machine learning techniques with graph-based characteristics to improve detection accuracy.

\subsubsection{Natural Language Processing (NLP)-Based Representation}
utilizes sophisticated NLP techniques to extract semantic information from documentation and code comments. 
The precise syntactic structures that are inherent to programming languages may not always be followed by NLP approaches, which are widely recognized for their effectiveness in capturing semantic patterns and purpose~\cite{mikolov2013efficient}.
However, these methods are just as good at deciphering semantic material as they are at processing natural languages, especially when it comes to code functionality and intent. 
While generally less resource-intensive than graph-based methods, NLP techniques still require significant processing, especially for semantic analysis and natural language processing tasks.
In~\cite{VulBERTa}, the authors leveraged a custom tokenization pipeline that combines the Byte-Pair Encoding algorithm with novel predefined code tokens to train an transformer based NLP model RoBERTa~\cite{liu1907roberta}(introduced by Facebook AI in 2019). 



\subsubsection{Embedding-Based Representation}
encodes the source code into low-dimens\\ional vector spaces using representations based on embeddings, which carefully maintain semantic links. 
These techniques usually provide a succinct numerical depiction of the code, identifying patterns useful for deducing syntax and semantics. 
V1Scan~\cite{V1Scan} effectively leveraged embedding-based representations to unveil vulnerabilities, thereby exemplifying the robustness of this approach in pinpointing security vulnerabilities.
Russell et al. ~\cite{Russell2018Draper} initially convert the tokens comprising the lexed functions into a fixed k-dimensional representation constrained to the range $[-1, 1]$. 
This representation is learned during the classification training through back-propagation applied to a linear transformation of a one-hot embedding.  
%


\tcbset{colframe=black, colback=white, boxrule=0.4mm, arc=4mm, auto outer arc, width=\textwidth}
\begin{tcolorbox}[width=\linewidth]
\textbf{Challenge:} \textit{Scalability remains a critical challenge, as many solutions struggle to efficiently handle large-scale codebases or real-time data due to computational limitations. Another significant issue lies in aligning deep learning models with suitable representation techniques, given their reliance on specific data formats and varying sensitivities to feature types, making effective matching essential for optimal performance. Furthermore, generalizing these methods across programming languages and development environments proves challenging, as approaches that perform well in one language may falter in others due to differences in syntax and semantic structures.}
\end{tcolorbox}

In this study, we propose a novel benchmark for the comparative analysis of five code representation techniques, employing three key scoring metrics Syntactic Capture Score (\( S_{syntactic} \)), Semantic Capture Score (\( S_{semantic} \)), and Complexity Score (\( S_{complexity} \)). To the best of our knowledge this is the first known attempt to establish such a comparison framework.


Each scoring matrix value ranging from 0 to 3 is controlled by multiple decision criteria, and scores are formulated following the principle of Multiple Criteria Decision Analysis (MCDA) \cite{wikipedia_mcda} and using the Weighted Decision Matrix (WDM) \cite{airfocus_wdm}. 
Syntactic Capture Score (\( S_{syntactic} \)) computed as a weighted sum of three critical sub-dimensions: Structural Fidelity (\(SF\)), Granularity (\(G\)), and Adaptability (\(A\)), each of which was assigned an efficiency value ranging from 1 to 3. The weights represent the relative importance of each sub-dimension in preserving the syntactic information of source code.

\begin{equation}
S_{syntactic} = 0.5 \cdot SF + 0.3 \cdot G + 0.2 \cdot A
\end{equation}

Semantic Capture Score (\( S_{semantic} \)) reflects the ability of a technique to model the meaning and contextual relationships in the source code. It is calculated as the weighted sum of Control and Data Flow Modeling (\(CDF\)), Context Awareness (\(CA\)), and Scalability (\(S\)).

\begin{equation}
S_{semantic} = 0.4 \cdot CDF + 0.3 \cdot CA + 0.3 \cdot S
\end{equation}

Complexity Score (\( S_{complexity} \)) evaluates the computational and implementation challenges associated with a source code representation technique. It is derived from the weighted sum of Computational Overhead (\(CO\)), Ease of Implementation (\(EI\)), and Scalability (\(S\)).

\begin{equation}
S_{complexity} = 0.5 \cdot CO + 0.3 \cdot EI + 0.2 \cdot S
\end{equation}

\begin{figure}[t] 
    \centering
    \includegraphics[scale=0.5]{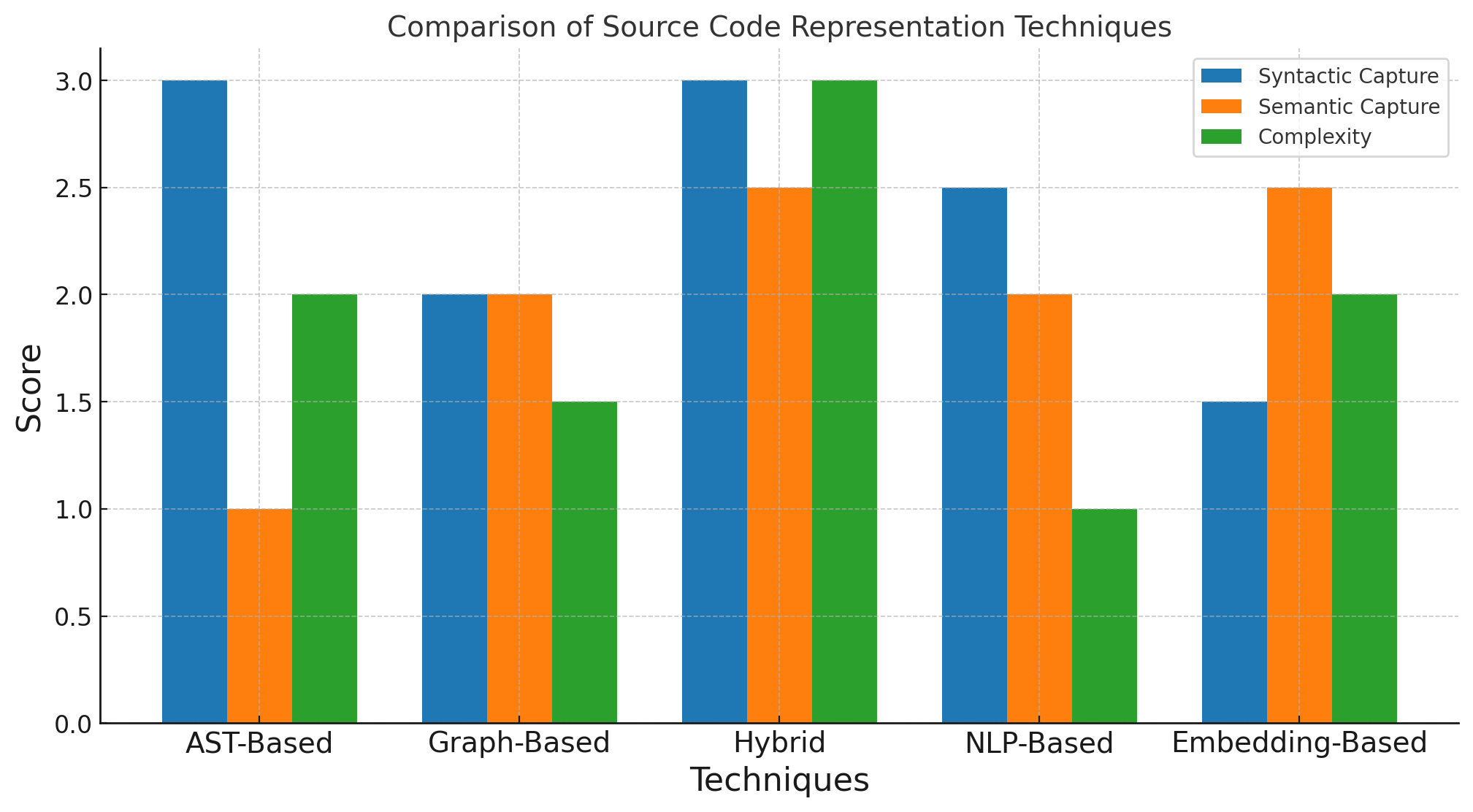} 
    \caption{Comparison of code representation techniques.}
    \label{fig:Code Representation}
    \vspace{-2em}
\end{figure}

Fig.~\ref{fig:Code Representation} presents the comprehensive comparison. Our scoring methodology demonstrates that the AST-Based technique achieves the highest score for syntactic capture but performs poorly in semantic capture and complexity. The Graph-Based technique demonstrates moderate performance across all three dimensions, reflecting a balanced but unremarkable profile. The Hybrid technique stands out with consistently high scores across all criteria, highlighting its robustness and versatility. In contrast, the NLP-Based technique excels in semantic capture but underperforms in syntactic capture and complexity. 
This scoring methodology ensures a robust and evidence-based evaluation of the techniques' effectiveness for source code representation and facilitates determining the appropriate technique to build a robust vulnerability detection model. 

\subsection{Model Design}

Upon completion of the pre-processing and code representations, the next critical step involves the vectorization and selection of deep learning models for training. The selection of a model impacts the choice of representation technique, as different models require data in specific formats and have varying sensitivity to feature granularity.This necessitates a careful matching process to ensure that the chosen technique aligns with the model's capabilities to achieve effective learning outcomes. Current state-of-the-art studies predominantly employ four vectorization techniques: One-hot encoding, Word2Vec (encompassing Continuous Bag of Words (CBOW) and Skip-gram)~\cite{mikolov2013efficient}, and Doc2Vec (including Paragraph Vector-Distributed Memory (PV-DM) and Paragraph Vector-Distributed Bag of Words (PV-DBOW))~\cite{le2014distributed}. Table ~\ref{table:dl_models} provides a comprehensive comparison of the representation types, models, and their specific characteristics utilized in state-of-the-art studies. The deep learning models explored in this study are categorized into three primary types: Sequence-based models, Graph-based models, and Hybrid models, which are described in the subsequent sections.

\begin{table*}[t]
\caption{Overview of Deep Learning Models for Vulnerability Detection}
\label{table:dl_models}
\centering
\tiny 
\setlength{\tabcolsep}{5pt} 
\begin{adjustbox}{max width=\textwidth} 
\begin{tabular}{>{\centering}p{1.4cm} >{\centering}p{0.7cm} >{\centering}p{0.8cm} >{\centering}p{2.5cm} >{\centering}p{1.3cm} >{\centering}p{2.3cm} p{2.3cm}}
\toprule
\textbf{Study} & \textbf{Type} & \textbf{Model} & \textbf{Components} & \textbf{Imple.} & \textbf{Features} & \makecell[c]{\textbf{Challenges}} \\
\midrule
\makecell{VulDeeP.\\~\cite{Li2018VulDeePecker}} & Seq. & BLSTM & \makecell{Long-range \\dependencies \\extraction} & \makecell{TensorFlow, \\Python\\} & \makecell{Improved \\vulnerability \\ identification} & \makecell{Limited to \\binary \\classification} \\
\midrule
\makecell{SySeVR\\~\cite{Li2021SySeVR}} & Seq. & BGRU & \makecell{Multiple neural \\networks, enhanced \\sequential pattern \\analysis} & \makecell{PyTorch, \\Python} & \makecell{Superior to \\shallow \\learning models} & \makecell{High \\computational \\ complexity} \\
\midrule
\makecell{\textmu VulDeeP.\\~\cite{Zou2019muvuldeppecker}} & Seq. & BLSTM & \makecell{Global-feature, \\local-feature, \\ feature-fusion \\models} & \makecell{TensorFlow, \\Keras} & \makecell{Multi-class \\ vulnerability \\ detection} & \makecell{Complex model \\ architecture} \\
\midrule
\makecell{POSTER\\~\cite{lin2017poster}} & Seq. & BLSTM & \makecell{AST-based \\function-level \\feature \\extraction} & \makecell{TensorFlow, \\Python} & \makecell{Cross-project \\ domain feature \\ extraction} & \makecell{Requires \\ comprehensive \\ AST construction} \\
\midrule
\makecell{Devign\\~\cite{zhou2019devign}} & Graph & GNN & \makecell{Data and \\ control dependency \\ code graphs,\\ Conv module} & \makecell{PyTorch, \\DGL} & \makecell{Extracts \\interesting \\ features from \\ source code} & \makecell{Handling large \\ graphs and \\ data dependencies} \\
\midrule
\makecell{DeepWukong\\~\cite{Cheng2021DeepWukong}} & Graph & GNN & \makecell{Compact, \\low-dimensional \\representation} & \makecell{PyTorch \\Geometric} & \makecell{Detects ten \\ different types of \\ vulnerabilities} & \makecell{Requires extensive \\ feature engineering} \\
\midrule
\makecell{DeepTective\\~\cite{zhang2020deeptective} }& Graph & \makecell{GCN$+$\\GRU} & \makecell{Combines GRU \\and GCN} & \makecell{TensorFlow, \\Keras} & \makecell{SQLi, XSS, \\and command \\ injection in PHP} & \makecell{Balancing GRU \\ and GCN training} \\
\midrule
\makecell{DRAPER\\~\cite{Russell2018Draper}} & Hybrid & \makecell{CNN$+$\\RNN} & \makecell{Feature extraction \\ from \\embedding-based \\ representations} & \makecell{TensorFlow,\\ Scikit-learn} & \makecell{Ensemble \\ classifiers for \\ improved results} & \makecell{Integration of \\multiple neural \\network types} \\
\bottomrule
\end{tabular}
\end{adjustbox}
\end{table*}



\subsubsection{Sequence-based Models} excel in managing sequential data due to their ability to identify temporal connections and patterns.
In domains like time series analysis and natural language processing (NLP), where temporal or sequential context is critical, these models exhibit remarkable performance. These models are especially useful for identifying intricate patterns that may be signs of security flaws because source code is sequential, much like the structure of natural language. This method not only increases vulnerability detection accuracy but also enables it to find subtle, context-dependent problems in the code.


Recurrent neural networks (RNN), Long Short-Term Memory networks (LSTM), Bidirectional Long Short-Term Memory (BLSTM), Gated Recurrent Units (GRU), Bidirectional Gated Recurrent Units (BiGRU), and Transformers are notable designs utilized in this field of vulnerability detection.
VulDeePecker~\cite{Li2018VulDeePecker} is an innovative pioneering study that employs a Bidirectional Long Short-Term Memory (BLSTM) paradigm to find vulnerabilities. Effective vulnerability identification is made possible by the BLSTM architecture, which is used to extract and understand long-range dependencies from code sequences. Several limitations of this study have been addressed by SySeVR~\cite{Li2021SySeVR}, a sophisticated system that employs a variety of neural networks to detect different types of vulnerabilities. In particular, the implementation of Bidirectional Gated Recurrent Units (BGRUs) has demonstrated superior effectiveness compared to shallow learning models. By leveraging BGRUs, SySeVR can more accurately capture and analyze complex sequential patterns in source code, thereby significantly enhancing the detection and identification of vulnerabilities. By altering the Bidirectional Long Short-Term Memory (BLSTM) architecture to enable multi-class vulnerability identification, {\textmu}Vuldeepecker ~\cite{Zou2019muvuldeppecker} builds on the work of VulDeePecker ~\cite{Li2018VulDeePecker}. Three different BLSTM models make up this sophisticated network: a feature-fusion model, a local feature-learning model, and a global feature-learning model. The global-feature learning model uses deep BLSTM layers and a preprocessing layer to extract global features. The deep BLSTM layers concentrate on extracting global features from the preprocessed data, whereas the preprocessing layer is mostly in charge of eliminating zero vectors.



Another notable bidirectional Long Short-Term Memory (BLSTM) model was proposed by Guanjun et al.~\cite{lin2017poster} for function-level vulnerability detection across cross-project domains. 
This model leverages the BLSTM architecture to extract features from functions represented using the Abstract Syntax Tree (AST) based representation method.

\subsubsection{Graph-based Models} are highly effective at handling complex relational data due to their ability to capture and utilize the interconnections and hierarchical structures within graph-structured data. These approaches encode complex relationships efficiently by representing data as graphs, making them well-suited for advanced analysis across various domains. Notable examples of these models include Graph Neural Networks (GNNs), Graph Convolutional Networks (GCNs), Graph Attention Networks (GATs), and Gated Recurrent Units (GRU). 
Zhou et al. proposed Devign~\cite{zhou2019devign} framework employs a Graph Neural Network (GNN) model to learn data and control dependency code graphs, incorporating a novel convolutional module designed to extract significant features from source code. Similarly, DeepWukong~\cite{cheng2021deepwokung} utilizes GNNs to embed code fragments into compact, low-dimensional representations, enabling the detection of ten distinct types of vulnerabilities. Extending beyond C/C++ applications, the DeepTective~\cite{zhang2020deeptective} system combines Gated Recurrent Unit (GRU) and Graph Convolutional Network (GCN) where the former operates on the linear sequence of source code tokens, and the latter operates on the Control Flow Graph (CFG) of the source code to identify vulnerabilities such as SQL injection (SQLi), cross-site scripting (XSS), command injection in PHP source code.



\subsubsection{Hybrid Models}, in an effort to maximize the advantages of sequential processing and structural comprehension, combine the best features of both sequence-based and graph-based methodologies. They offer a more comprehensive depiction of the data by merging these approaches, which may improve performance on a range of tasks, including Code Structures and Behaviors understanding, and feature-learning enhancements. Russel et al.~\cite{Russell2018Draper} conducted an in-depth exploration of both Convolutional Neural Networks (CNNs) and Recurrent Neural Networks (RNNs) for feature extraction from embedding-based representations of source code. Following the extraction of these features, they employed an ensemble of powerful classifiers, such as random forests and extremely randomized trees. This combination yielded the most effective results in their study, demonstrating the superior performance of these advanced machine learning techniques in vulnerability detection.



\tcbset{colframe=black, colback=white, boxrule=0.4mm, arc=4mm, auto outer arc, width=\textwidth}
\begin{tcolorbox}[width=\linewidth]
\textbf{Challenge:} \textit{Training models for vulnerability detection often incurs significant computational costs and prolonged inference times. This has prompted ongoing research into optimization techniques, model compression, and hardware acceleration to improve efficiency without compromising performance. Additionally, while some models excel on specific datasets and vulnerability classes, their ability to generalize to new, unseen vulnerabilities remains a persistent issue. Lastly, determining the optimal model for feature learning is a critical challenge, as different models demonstrate varying learning capacities when applied to the same dataset, complicating the selection process.}
\end{tcolorbox}

\subsection{Model Performance Evaluation}

Standardization in model performance evaluation helps by recognizing real progress in the field and enables researchers to compare new models with existing benchmarks. 
Additionally, comprehensive assessments involving multiple metrics can reveal the strengths and weaknesses of different models, leading to more reliable vulnerability detection systems.

We explore the metrics commonly used by existing works and draw a comparative analysis in Table ~\ref{tab:evaluation_metrics}, listing them as follows:

\noindent \textbf{Accuracy ($\boldsymbol{Acc}$)} is the proportion of accurately predicted instances compared to the total instances.

{
\tiny
\begin{equation}
Acc = \frac{TP + TN}{TP + TN + FP + FN}
\end{equation}
}

where $TP$, $TN$, $FP$, and $FN$ denote the true positive instances, true negative instances, false positive instances, and false negative instances, respectively.

\noindent \textbf{Precision ($\boldsymbol{Pre}$)} presents the ratio of correctly predicted positive instances to the total number of positive instances.

{
\tiny
\begin{equation}
Pre = \frac{TP}{TP + FP}
\end{equation}
}

\noindent \textbf{Recall ($\boldsymbol{Rec}$)} represents the ratio of true positive instances to the sum of true positive instances and false negative instances.

%

{\tiny
\begin{equation}
Rec = \frac{TP}{TP + FN}
\end{equation}}

\noindent \textbf{F1-score ($\boldsymbol{F1}$)} is the balanced measure that combines precision and recall, making it particularly useful for evaluating imbalanced datasets.

{\tiny
\begin{equation}
F1 = 2 \cdot \frac{Pre \cdot Rec}{Pre + Rec}.
\end{equation}
}

\noindent \textbf{False Positive Rate ($\boldsymbol{FPR}$)} is the proportion of negative instances that are incorrectly classified as positive, calculated as the ratio of false positives to the sum of false positives and true negatives.

{
\tiny
\begin{equation}
FPR = \frac{FP}{FP + TN}
\end{equation}
}

\noindent \textbf{False Negative Rate ($\boldsymbol{FNR}$)} represents the proportion of positive instances that are incorrectly classified as negative, which is especially important in situations where missing positive cases can have serious consequences.
{
\tiny
\begin{equation}
FNR = \frac{FN}{FN + TP}
\end{equation}
}

\tcbset{colframe=black, colback=white, boxrule=0.4mm, arc=4mm, auto outer arc, width=\textwidth}
\begin{tcolorbox}[width=\linewidth]
\textbf{Challenge:} \textit{A key challenge is reducing FPR and FNR to minimize the need for manual intervention, thereby enhancing the reliability and effectiveness of automated vulnerability detection systems.}
\end{tcolorbox}

\begin{table*}[t]
\centering
\caption{Performance Metrics of Various Vulnerability Detection Approaches}
\label{tab:evaluation_metrics}
\tiny 
\setlength{\tabcolsep}{7pt} 
\begin{adjustbox}{max width=\textwidth} 
\begin{tabular}{p{1.5cm} p{0.8cm} p{0.8cm} p{0.8cm} p{0.8cm} p{0.8cm} p{0.8cm} p{0.8cm} p{0.8cm} p{0.8cm}}
\toprule
\makecell[c]{\textbf{Work}} & \textbf{$Pre$} & \textbf{$Rec$} & \textbf{$Acc$} & \textbf{$F1$} & \textbf{$FPR$} & \textbf{$FNR$} & \textbf{$MCC$} & \textbf{$PR$-$AUC$} & \textbf{$ROC$-$AUC$} \\
\midrule
\makecell{SySeVR\\~\cite{Li2021SySeVR}} & 90.80\% & - & 98.00\% & 92.60\% & 1.40\% & 5.60\% & 90.50\% & - & - \\
\hline
\makecell{DeepWK\\~\cite{cheng2021deepwokung}} & - & - & 97.40\% & 95.60\% & - & - & - & - & - \\
\hline
\makecell{FUNDED\\~\cite{wang2020FUNDED}} & 92.00\% & 94.00\% & 92.00\% & 94.00\% & - & - & - & - & - \\
\hline
\makecell{DRAPER\\~\cite{Russell2018Draper}} & - & - & - & 82.40\% & - & - & 67.20\% & 91.60\% & 93.60\% \\
\hline
\makecell{VulDeeP.\\~\cite{Li2018VulDeePecker}} & 78.60\% & - & 83.10\% & 80.80\% & 22.90\% & 16.90\% & - & - & - \\
\hline
\makecell{{\textmu}VulDeeP.\\~\cite{Zou2019muvuldeppecker}} & - & - & - & 94.6\% & 0.02\% & 5.73\% & - & - & - \\
\hline
\makecell{Devign\\~\cite{zhou2019devign}} & - & - & 75.56\% & 27.25\% & - & - & - & - & - \\
\hline
\makecell{Subhan et al.\\~\cite{subhan2022deep}} & 92.00\% & 98.50\% & 91.19\% & 95.50\% & - & - & - & - & - \\
\bottomrule
\end{tabular}
\end{adjustbox}
\end{table*}

Reducing these rates significantly minimizes the need for manual intervention by security experts in automated vulnerability detection, thereby improving the dependability and efficiency of these systems, which is a key goal within the research community.

\noindent \textbf{Matthews Correlation Coefficient ($\boldsymbol{MCC}$)} measures the quality of binary classifications by assessing the correlation between the predicted and actual outcomes.
MCC is particularly useful for evaluating models on imbalanced datasets as it provides a more informative and truthful score in such scenarios.

{
\tiny
\begin{equation}
MCC = \frac{(TP \cdot TN) - (FP \cdot FN)}{\sqrt{(TP + FP)(TP + FN)(TN + FP)(TN + FN)}}
\end{equation}
}

\noindent \textbf{Precision-Recall Curve ($\boldsymbol{PR}$-$\boldsymbol{AUC}$)} plot the trade-off between $Pre$ and $Rec$ at various threshold values.

\noindent \textbf{Receiver Operating Characteristic ($\boldsymbol{ROC}$-$\boldsymbol{AUC}$)} presents the graphic that illustrates the trade-off between TPR and FPR at various threshold values. 
Better discriminatory ability is shown by greater values of the area under the ROC curve ($ROC$-$AUC$), which offers a single scalar value summarizing the model's performance.

\begin{figure*}[t] 
    \centering
    \includegraphics[width=0.99\textwidth]{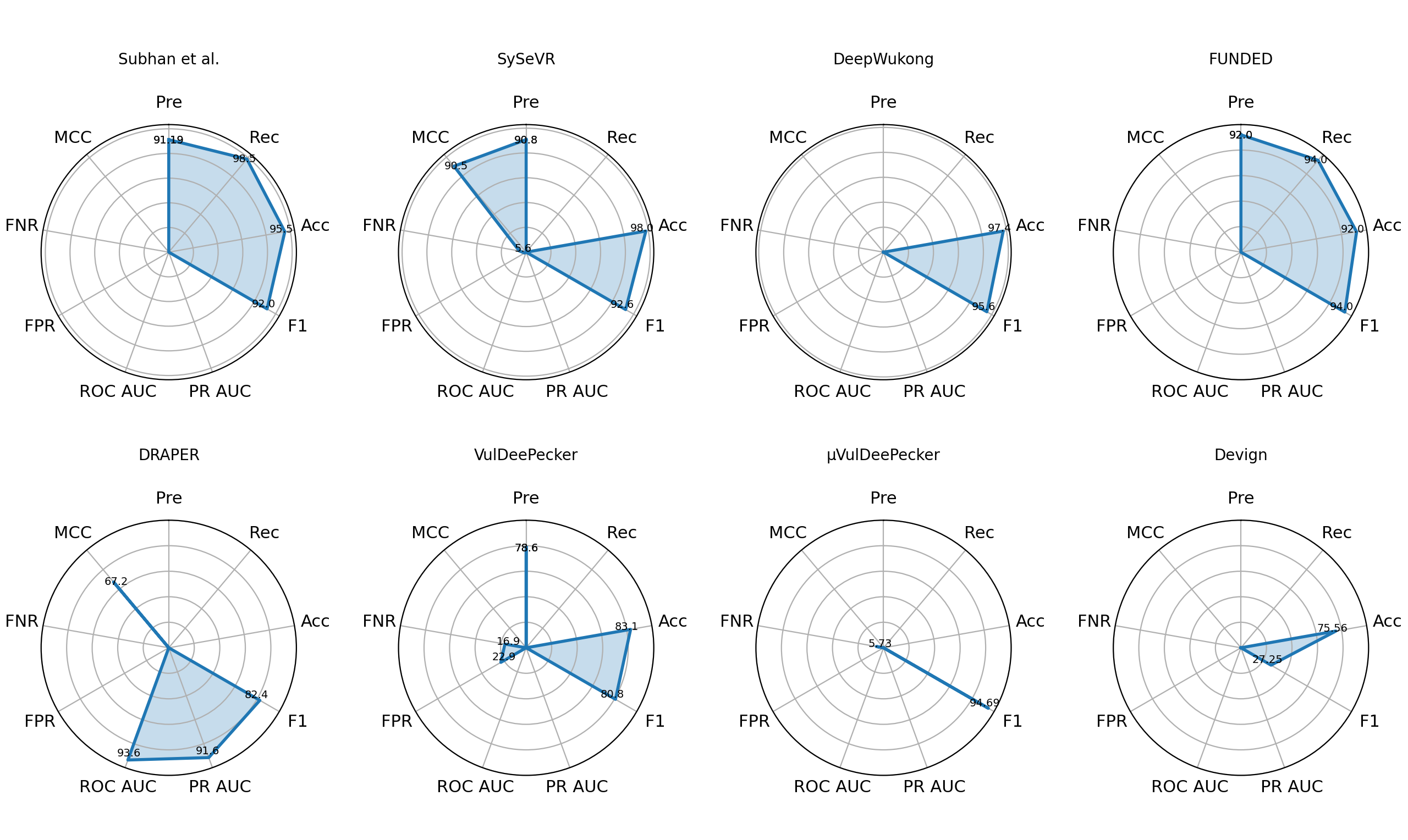} 
    \caption{\small Visualizing Performance Metrics of Vulnerability Detection Models} 
    \label{fig:evaluation}
    \vspace{-1em} 
\end{figure*}

Fig.~\ref{fig:evaluation} presents charts that visualize the evaluation metrics for various software vulnerability detection models. Each chart represents an individual model, depicting its performance across the examined range of metrics. 
This detailed visualization enables an intuitive comparison of the models' performance across different dimensions of vulnerability detection, thereby elucidating their strengths and identifying areas requiring enhancement.

\tcbset{colframe=black, colback=white, boxrule=0.4mm, arc=4mm, auto outer arc, width=\textwidth}
\begin{tcolorbox}[width=\linewidth]
\textbf{Challenge:} \textit{ Establish baseline models and dataset that represent the current SOTA. These models will serve as benchmarks against new models and datasets.}
\end{tcolorbox}



A comparison with a relevant and reasonable baseline is crucial to demonstrate the improvements provided by a particular model. However, each study often introduces its own baseline for comparison, highlighting the need for a domain-specific standardized baseline. This standardization would facilitate more consistent and meaningful evaluations across different studies, ensuring that improvements are accurately quantified and compared.

\tcbset{colframe=black, colback=white, boxrule=0.4mm, arc=4mm, auto outer arc, width=\textwidth}
\begin{tcolorbox}[width=\linewidth]
\textbf{Challenge:} \textit{Real-world project implementation is seldom practiced in studies, yet it should be established as a benchmark for evaluation.}
\end{tcolorbox}

\subsection{Real-world Project Implementation}
Deploying models in practical settings to identify both known and previously unknown vulnerabilities is essential and that completes our vulnerability detection life cycle. The practical usefulness and robustness of these models are shown through real-world testing, which also exposes the models' true performance outside of controlled experimental settings. Such an implementation provides essential feedback for further model improvement in addition to highlighting the models' strengths and limitations. Through the examination of real-world problems, scientists can guarantee that their models are both theoretically and practically solid, resulting in automated vulnerability detection systems that are more dependable and efficient. This strategy ultimately propels the field's progress by bridging the gap between scholarly study and real-world application. Table~\ref{tab:perform} presents the real-world implementation of a selected set of studies.


A significant challenge is ensuring that vulnerability detection models, which perform well in controlled experimental settings, can also demonstrate practical feasibility and effectiveness in real-world applications.

\tcbset{colframe=black, colback=white, boxrule=0.4mm, arc=4mm, auto outer arc, width=\textwidth}
\begin{tcolorbox}[width=\linewidth]
\textbf{Challenge:} \textit{Developed models capable of identifying known and previously unknown vulnerabilities when deployed in real-world environments.}
\end{tcolorbox}

SySeVR~\cite{Li2021SySeVR} was implemented in real-world scenarios involving four software products (Libav, Seamonkey, Thunderbird and Xen) and reported 15 vulnerabilities not listed in the National Vulnerability Database (NVD), 7 of which were previously unknown vulnerabilities are subsequently patched by their respective vendors. The remaining 8 vulnerabilities had already been addressed by the vendors in newer versions of the software products. This highlights the crucial capability of models to detect not only known vulnerabilities but also previously unknown threats when deployed in real-world software environments—an essential attribute for effective vulnerability detection systems. For the real-world implementation, VulDeePecker~\cite{Li2018VulDeePecker} analyzed 20 versions of the three software products (Xen, Seamonkey, and Libav) by identifying 4 vulnerabilities later fixed by the vendor. 
Implementation on three real-world software products detected 16 vulnerabilities, 14 of which corresponded to known vulnerability patterns from
{\textmu}VulDeePecker~\cite{Zou2019muvuldeppecker}. In general, evaluating a model's actual efficacy in a range of real-world circumstances can be difficult due to differences in vendors' mitigation of vulnerabilities found in their software products. It is cumbersome to collect and apply real-world input to the models in an efficient manner since this involves constant assessment and modification based on real-world use cases.

\tcbset{colframe=black, colback=white, boxrule=0.4mm, arc=4mm, auto outer arc, width=\textwidth}
\begin{tcolorbox}[width=\linewidth]
\textbf{Challenge:} \textit{Variations in vendor practices for handling vulnerabilities and the need for ongoing feedback from real-world deployments present challenges in accurately assessing and refining the effectiveness of vulnerability detection models.}
\end{tcolorbox}

\begin{table*}[t]
\caption{Summary of Studies on Vulnerability Detection}\label{tab:perform}
\centering
\tiny 
\setlength{\tabcolsep}{6pt} 
\begin{adjustbox}{max width=\textwidth} 
\begin{tabular}{p{1.2cm} p{0.8cm} p{0.4cm} p{1.3cm} p{1.5cm} p{2cm} p{2.5cm} p{2cm}}
\toprule
\makecell{\textbf{Dataset}}  & \textbf{Pos.:Neg.} & \textbf{Gran.} & \makecell{\textbf{Repr.}} & \textbf{DL Model} & \textbf{Exp. Setting} & \makecell{\textbf{Performance}} & \makecell{\textbf{Project impl.}} \\
\midrule
\makecell{SySeVR\\~\cite{Li2021SySeVR}} & 13:87 & $L3$ & \makecell{Syntax, \\Semantics, \\Vector} & \makecell{Bi-GRU} & \makecell{Python, \\TensorFlow, \\NVIDIA GPU} & \makecell{FPR 1.4\%, FNR 5.6\%, \\Acc 98\%, Pre 90.8\%, \\F1 92.6\%, MCC 90.5\%} & \makecell{Detected 15 \\vulnerabilities \\in 4 products} \\
\midrule
\makecell{DeepWK.\\~\cite{cheng2021deepwokung}}  & 28:72 & $L3$ & Slice-graph & \makecell{GNN} & \makecell{PyTorch \\Geometric} & \makecell{Acc 97.4\%, F1 95.6\%} & -\\
\midrule
\makecell{FUNDED\\~\cite{wang2020FUNDED}}  & 50:50 & $L2$ & \makecell{AST} & \makecell{GGNN+GRU} & \makecell{TensorFlow, \\ Python  \\scikit-learn} & \makecell{Acc 92\%, Pre 92\%, \\Rec 94\%, F1 94\%} & - \\
\midrule
\makecell{Draper\\~\cite{Russell2018Draper}}  & 6:94 & $L2$ & \makecell{Embedded} & \makecell{CNN+RNN} & \makecell{N/A} & \makecell{PR-AUC 92\%, \\ ROC-AUC 94\%, \\ MCC 67\%, F1 82\%} & - \\
\midrule
\makecell{VulDeeP.\\~\cite{Li2018VulDeePecker}} & 29:71 & $L3$ & \makecell{Embedded} & \makecell{BLSTM} & \makecell{Python,  \\ Theano, \\ Keras} & \makecell{Rec 83.1\%, Pre 78.6\%, \\ F1 80.8\%} & \makecell{Detected 4 \\vulnerabilities \\in 3 products} \\
\midrule \makecell{\textmu VulDeeP.\\~\cite{Zou2019muvuldeppecker}} &  24:76 & $L3$ & \makecell{Graph} & \makecell{BLSTM} & \makecell{Intel CPU, \\NVIDIA GPU, \\Linux OS} & \makecell{FPR 0.02\%, \\ FNR 5.73\%, F1 94.22\%} & \makecell{Detected 16 \\vulnerabilities \\in 3 products} \\
\midrule
\makecell{Devign\\~\cite{zhou2019devign}} &  45:55 & $L2$ & \makecell{Graph} & \makecell{GGRN} & \makecell{NVIDIA Tesla \\ M40/P40} & Acc 75.56\%, F1 27.25\% & - \\
\midrule
\makecell{SySvr,\\
CMD\\~\cite{subhan2022deep}} & 13:87 & $L3$ & \makecell{Graph} & \makecell{CNN, LSTM, \\ GRU \\ CNN-LSTM} & \makecell{Python, \\ Keras, \\ TensorFlow2} & \makecell{Pre 92\%, Rec 99\%, \\ Acc 91\%, F1 95\%} & - \\
\midrule
\makecell{SARD\\~\cite{hussain2024vulnerability}} &  31:69 & L4 & \makecell{Graph} & \makecell{QCNN} & \makecell{Python, \\Keras, \\TensorFlow2} & \makecell{Pre 97\%, Rec 98\%, \\ Acc 99\%, F1 97\%} & - \\
\bottomrule
\end{tabular}
\end{adjustbox}
\end{table*}

\section{Future Research Direction Discussion}
\label{sec:Future Research}

Our study provides important insights and avenues for future investigation, laying a solid platform for future research in this area. Addressing challenges and creating workable answers to them would constitute a substantial research contribution and advance the study. In order to guarantee that vulnerability detection models are trained and assessed on a variety of pertinent data, future research should place a high priority on the development of approaches for choosing and curating datasets that accurately represent real-world scenarios. In order to balance accuracy and computational performance, this entails determining the proper degree of granularity in detection, which enables models to identify vulnerabilities efficiently while reducing false positives and negatives. The optimization of code representations is also crucial, as is tackling the difficulties in generalizing these representations across various programming languages and contexts. This optimization should strike a balance between computational practicality and detailed syntactic and semantic capture. To ensure consistent and meaningful performance comparisons across different models, robust baselines and standardized evaluation metrics should be developed. Multiple metrics and visualization techniques should be used to better capture model performance, especially in the context of imbalanced datasets. Furthermore, developing deep learning models' interpretability and explainability will be essential to fostering transparency and confidence in real-world uses. Finally, research should focus on creating standardized frameworks for evaluating models in diverse real-world software environments, accounting for variations in how vendors handle and patch vulnerabilities, and systematically gathering feedback from real-world deployments to continuously refine and improve model performance and robustness.

\section{Conclusion}

\label{sec:conclusion}


 In this survey, we meticulously outlined the various phases involved in applying deep learning techniques to detect vulnerabilities in source code. By identifying and selecting seminal and highly impactful studies from the past seven years, we have highlighted their influence on subsequent research in the domain of deep learning-based vulnerability detection. For each component of the defined life cycle, we addressed key aspects and challenges within each phase. Our examination of the methodologies employed by the selected studies revealed significant challenges encountered throughout the life cycle. Moreover, we pinpointed existing research gaps stemming from these challenges, providing a comprehensive overview of current and future directions in the field. In conclusion, this survey not only highlights the progress made in applying deep learning to vulnerability detection but also emphasizes the need for continued exploration and development in this critical area of research. Future work should focus on overcoming identified challenges, exploring novel methodologies, and fostering collaboration between academia and industry to achieve meaningful advancements in vulnerability detection.

\bibliographystyle{splncs04}
\small


\section*{Authors}
\noindent {\bf Md Nizam Uddin} obtained a Bachelor’s degree in Computer Science from the Islamic University of Technology (IUT). He spent seven years as a Software Quality Assurance Engineer at Therap Services LLC, where he specialized in software security, vulnerability assessment, and compliance verification. Currently, he is pursuing a Ph.D. in Computer Science at the University of Louisiana at Lafayette, focusing on cutting-edge research in cybersecurity and industrial control system security. His research interests span information and application security, security challenges in cyber-physical systems, and the resilience of critical infrastructure.\\

\noindent {\bf Yihe Zhang} received M.S. degree from SYSU-CMU Joint Institute of Engineering, Sun-Yat Sen University and Carnegie Mellon University, in 2016. He is currently working toward the Ph.D. degree in the School of Computing and Informatics, University of Louisiana at Lafayette. His research interests include machine learning applications, knowledge graph and data-driven security. He received the Best Paper Award and the Distinguished Paper Award from DSN 2023.\\

\noindent {\bf Dr. Xiali Hei} is an Alfred and Helen M. Lamson Endowed associate professor in the School of Computing and Informatics at the University of Louisiana at Lafayette since August 15th, 2023. She was an Alfred and Helen M. Lamson Endowed assistant professor in the School of Computing and Informatics at the University of Louisiana at Lafayette from August 2017 to August  15th, 2023. Prior to joining the University of Louisiana at Lafayette, she was an assistant professor at Delaware State University from 2015-2017 and an assistant professor at Frostburg State University from 2014-2015. She was awarded Alfred and Helen M. Lamson Endowed Professorship, an Outstanding Achievement Award in Externally Funded Research, five NSF awards, a Meta Research award, IEEE SP 2024 Distinguished paper award, EAI SmartSP 2023 Best paper award, etc.  Her papers were published at IEEE S\&P, USENIX Security Symp., ACM CCS, IEEE INFOCOM, IEEE Euro S\&P, RAID, ASIACCS, etc. She is a PC member of the USENIX Security Symp., IEEE EuroS\&P, PST, IEEE GLOBECOM, SafeThings, VehivleSec, IEEE ICC, WASA, etc. She was the general chair of EAI SmartSP 2024.  Her research interests are Fast Encryption, Security of Wireless Medical Devices, Biometrics security, Mobile security, Device Security, Network Forensics. \\

\end{document}